\documentclass{article}
\usepackage{spconf,amsmath,graphicx}

\usepackage{enumitem}
\setlist{nosep, leftmargin=14pt}
\usepackage{placeins}
\usepackage{mwe} % to get dummy images
\usepackage{multirow}
\usepackage{xcolor}

\title{Multimodal Deep Learning to Differentiate Tumor Recurrence from Treatment Effect in Human Glioblastoma}
\name{
\parbox{\linewidth}{
\centering
Tonmoy Hossain* $^1$, Zoraiz Qureshi* $^{1,2}$, Nivetha Jayakumar $^{1,2}$, Thomas Eluvathingal Muttikkal $^2$, \\ \textit{Sohil Patel $^2$, David Schiff $ ^3$, Miaomiao Zhang $^{1,4}$ and Bijoy Kundu $^{2,5}$}
\thanks{* Equal contributing authors.}}}
\address{
\parbox{\linewidth}{
\centering
Departments of Computer Science$^1$, Radiology and Medical Imaging$^2$, Neurology$^3$, Electrical\\ and Computer Engineering$^4$, and Biomedical Engineering$^5$, University of Virginia, USA\\}}
\begin{document}
\maketitle
\begin{abstract}
Differentiating tumor progression (TP) from treatment-related necrosis (TN) is critical for clinical management decisions in glioblastoma (GBM). Dynamic FDG PET (dPET), an advance from traditional static FDG PET, may prove advantageous in clinical staging. dPET includes novel methods of a model-corrected blood input function that accounts for partial volume averaging to compute parametric maps that reveal kinetic information. In a preliminary study, a convolution neural network (CNN) was trained to predict classification accuracy between TP and TN for $35$ brain tumors from $26$ subjects in the PET-MR image space. 3D parametric PET Ki (from dPET), traditional static PET standardized uptake values (SUV), and also the brain tumor MR voxels formed the input for the CNN. The average test accuracy across all leave-one-out cross-validation iterations adjusting for class weights was $0.56$ using only the MR, $0.65$ using only the SUV, and $0.71$ using only the Ki voxels. Combining SUV and MR voxels increased the test accuracy to $0.62$. On the other hand, MR and Ki voxels increased the test accuracy to $0.74$. Thus, dPET features alone or with MR features in deep learning models would enhance prediction accuracy in differentiating TP vs TN in GBM.
\end{abstract}
\begin{keywords}
Dynamic FDG Brain PET, Model Blood Input, MRI, 3D CNN, Tumor Classification
\end{keywords}
\section{Introduction}
Glioblastoma (GBM) is a highly aggressive brain neoplasm with a median survival of $15$ months. Surgical resection and adjuvant chemo-radio therapy are palliative treatments alone~\cite{stupp2005radiotherapy}. The latter induces changes to brain tissue which in-turn produces similar neuroimaging changes to tumor recurrence~\cite{kim2010differentiating}, hence making it critical for clinical management decisions to differentiate between recurring tumor (TP) and treatment effect (TN). Positron emission tomography (PET) with Fluorine-18 fluorodeoxyglucose (18F-FDG) as a surrogate marker for glucose metabolism, represents an imaging technique that can provide pathophysiologic and diagnostic data in this clinical setting. The current standard of care regarding clinical FDG PET is a qualitative and visual analysis by performing comparisons to the contra-lateral and other brain regions. Static standardized PET uptake values (SUV) measured at a specific time point post-FDG injection have been widely used as a semi-quantitative measure \cite{nozawa2013glucose}. However, SUV does not reliably differentiate tumor from therapy effect, as it can depend on several factors such as body weight and blood glucose level.  Dynamic FDG PET (dPET), an advance from traditional static FDG PET, may prove advantageous in clinical staging. dPET includes novel methods of a model-corrected blood input function that accounts for partial volume averaging to compute a parametric rate of uptake, or Ki, maps that reveal kinetic information.

In this work, we propose a multi-modal tumor classification framework consisting of a dual-encoder 3D convolutional neural network (CNN) and found that the metabolic information from dPET assists MRI in classification and the CNN combined with multiple image modalities performs better in differentiating tumor progression from treatment effect in human GBM.

 %capture the standard of care, magnetic resonance imaging (MRI) and PET features. We addressed--- metabolic information in dynamic PET assists MRI in the classification between radiation necrosis and tumor progression, and (II) the combination of image modalities with dual-encoder CNN performs better than a dual-channel-based CNN.

\begin{figure*}[h]
    \centering
    \includegraphics[scale=0.5]{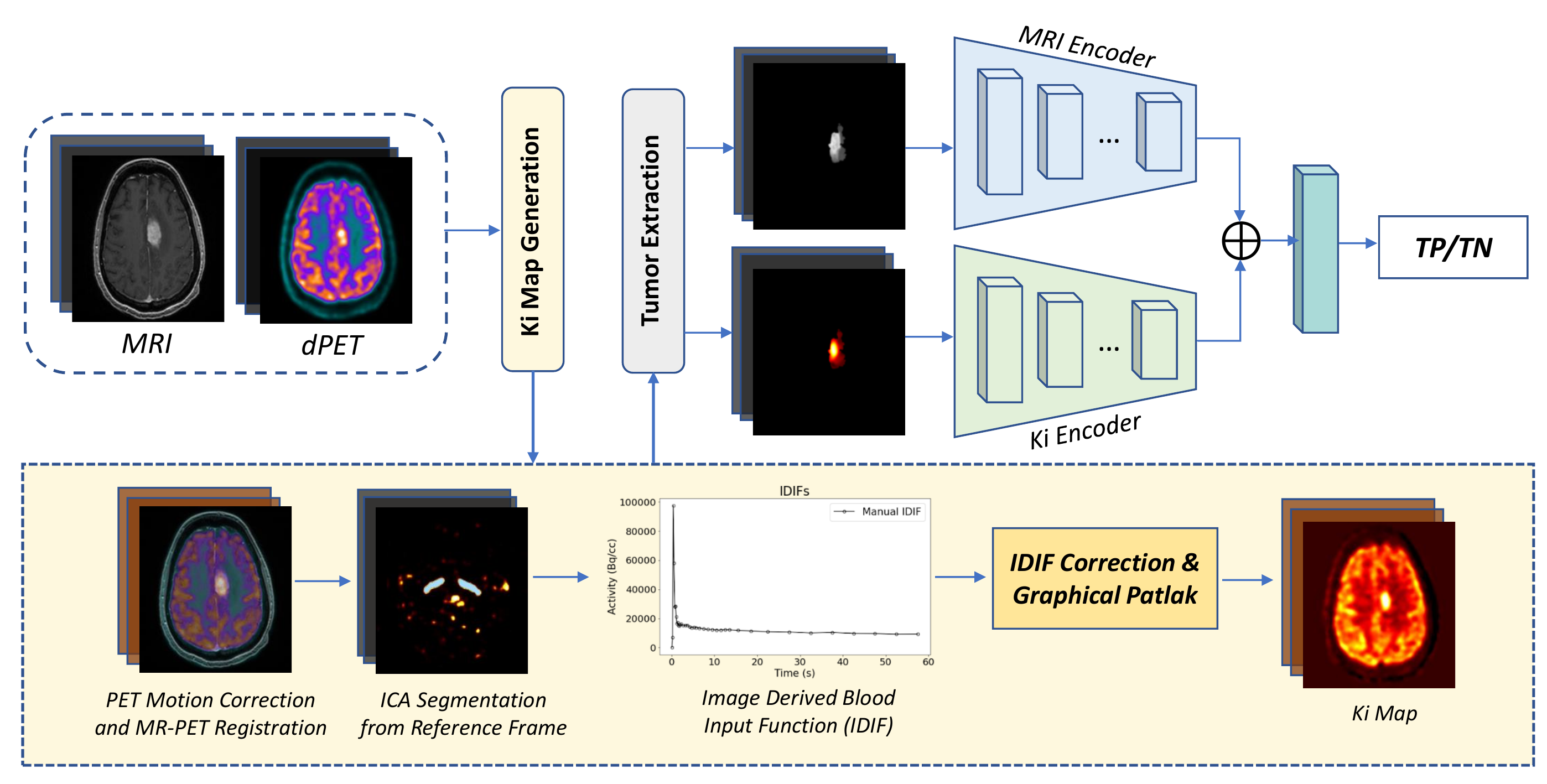}
    \caption{Proposed multi-modal classification network and data processing pipeline.}
    \label{fig:prop_model}
\end{figure*}

\section{Proposed Multi-modal Classification}
In this section, we present our proposed multi-modal classification framework and its components. Our model consists of two modules -- (I) Ki Map Generation \& Tumor Extraction and (II) Tumor Classification shown in \textbf{Fig. \ref{fig:prop_model}}. Details of our proposed methodology are introduced as follows.  

\vspace{-5pt}

\subsection{Ki Map Generation}
\textbf{PET Motion Correction and MR-PET Registration.} To correct patient head motion during the dynamic PET scan, we perform motion correction and co-registration to MR space as described \cite{seshadri2021dynamic}. Since static PET is the standard of care for clinical diagnosis, we also compute Standardized Uptake Value (SUV) maps for evaluation.

\textbf{ICA segmentation and IDIF derivation.} To compute parametric glucose rate-of-uptake or Ki maps from co-registered dynamic PET scans representing kinetic information, we compute an image-derived blood input function (IDIF) from the ICA as the region of interest. An early reference frame is selected for semi-automated volumetric annotation of the ICA using thresholding and islanding in 3D Slicer. The resultant ICA segmentation is convolved to compute the average blood time-activity curve across all time frames to produce the tracer time-activity curve serving as the IDIF.
 
\textbf{IDIF correction and graphical Patlak for parametric Ki computation.} To generate the parametric brain PET maps, a model-corrected blood input function (MCIF) is computed by optimizing the image-derived input function (IDIF) derived from the ICA as described \cite{seshadri2021dynamic}, to account for partial volume recovery of the blood input.

Finally, the computed MCIF and whole-brain PET data are fed into a graphical Patlak model \cite{patlak1983graphical}. The model performs a voxel-wise linear regression on the data to derive the rate of FDG uptake, Ki, as the slope. By analyzing millions of voxels across the entire PET volume, a parametric 3D Ki map is computed. 

\subsection{Tumor Extraction}
To specifically ensure that the model only focuses on tumor voxels instead of brain background voxels, we perform a semi-automated segmentation of tumor volumes using the seed-based region growing segmentation tool in 3D Slicer. This is followed by Gaussian smoothing to get a conservative mask of the abnormality and then verification from clinical experts. These masks are dropped onto co-registered PET Ki, SUV, and T1-weighted MRI images to mask out the tumor.   

\begin{figure}[htb!]
\begin{minipage}[b]{1.0\linewidth}
  \centering
  \centerline{\includegraphics[width=8.5cm]{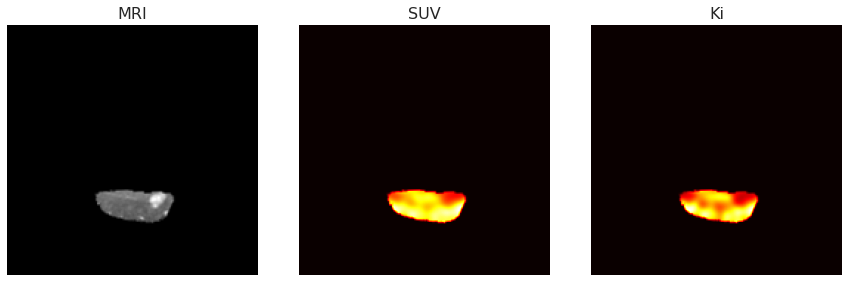}}
\end{minipage}
\caption{Examples processed tumor images from each modality (MR, SUV, Ki) used as network inputs.}
\label{fig:tumors}
\end{figure}

3D parametric PET Ki, static PET standardized uptake value (SUV), and MR tumor voxels are extracted in the same image space.  To perform the extraction of 3D tumor voxels while consistently preserving positional information, as a pre-processing step, the images and tumor mask for each subject are co-registered to the SRI24 atlas \cite{rohlfing2008sri24} with dimensions $(240, 240, 155)$ and re-oriented into Left Posterior Superior (LPS) orientation. This is done by registering the MR to the atlas after performing temporary $N4$ bias field correction and mutual information for rigid registration using the Cancer Imaging Phenomics Toolkit (CaPTk) \cite{davatzikos2018cancer}, then using the computed transformation to bring the other modalities to the same space after which the mask is applied. The masked images were further center cropped to the brain to $(170, 170, 120)$ dimensions by computing a minimum viable bounding box across all datasets to remove unnecessary background voxels and bring focus to the tumors. In \textbf{Fig. \ref{fig:tumors}}, we visualize the extracted tumor voxels for each modality. We observe a higher signal-to-noise ratio showcasing distinct tumor metabolic features within the Ki maps compared to SUVs.
\begin{table}[htb]
\centering
\caption{Classification performance metrics for different image modalities and network architectures.\\}
\begin{tabular}{|l|c|c|c|c|}
\hline
\textbf{Model}                                                             & \textbf{Modality} & \textbf{Accuracy} & \textbf{Precision} & \textbf{Recall} \\ \hline
\multirow{3}{*}{\begin{tabular}[c]{@{}l@{}}Single-\\ encoder\end{tabular}} & MR                & 0.56              & 0.71               & 0.68            \\ \cline{2-5} 
                                                                           & SUV               & 0.65              & 0.78               & 0.72            \\ \cline{2-5} 
                                                                           & Ki                & 0.71              & 0.80               & 0.80            \\ \hline
\multirow{3}{*}{\begin{tabular}[c]{@{}l@{}}Dual-\\ channel\end{tabular}}   & MR+SUV            & 0.53              & 0.72               & 0.60            \\ \cline{2-5} 
                                                                           & Ki+SUV            & 0.71              & 0.78               & 0.84            \\ \cline{2-5} 
                                                                           & MR+Ki             & 0.71              & 0.80               & 0.65            \\ \hline
\multirow{3}{*}{\begin{tabular}[c]{@{}l@{}}\textbf{Dual-}\\ \textbf{encoder}\end{tabular}}   & MR+SUV            & 0.62              & 0.75               & 0.72            \\ \cline{2-5} 
                                                                           & Ki+SUV            & 0.65              & 0.76               & 0.76            \\ \cline{2-5} 
                                                                           & \textbf{MR+Ki}             & \textbf{0.74}              & \textbf{0.80}               & \textbf{0.84}            \\ \hline
\end{tabular}
\label{table:comp}
\end{table}

\vspace{-10pt}

\subsection{Multi-modal Architecture}
After extracting tumor voxels from different image modalities, we develop a dual-encoder CNN architecture for multi-modal classification. The average dimensions of the tumors are relatively much smaller than the bounding brain region. For these low-dimensional tumor volumes and due to the low number of samples, there is a high probability that employing any SOTA classification network (has more than $30$M parameters e.g. ResNet18: $\sim34$M, VGGNet16: $\sim138$M) will overfit. Hence, we develop a custom convolutional neural network with a limited number of encoder layers. For each modality, we utilize a shallow 3D convolutional encoder architecture with $3$ convolutional layers (kernel size $3$, and a number of filters $B\times2$, $B\times4$ and $B\times4$ where the number of base filters $B$ is empirically set to $8$). The output latent feature vectors from the encoder of each modality are flattened and fused by concatenation before being fed into the fully connected layers. Additionally, a drop-out layer is added after the first dense layer with a rate of $0.2$ for regularization.

\vspace{-5pt}

\section{Experimental Evaluation}

\subsection{Dataset}
Our dataset consists of dynamic 18F-FDG PET scans for $26$ subjects with GBM obtained using the whole-body time of flight (TOF) Siemens Biograph mCT scanner with attenuation correction over time. Dynamic acquisition consisted of an intravenous $\sim10$ mCi tracer injection over $10$ seconds with the initiation of a $60$-minute scan in list-mode format. T1-weighted MPRAGE MRI scans are also obtained for each subject using the Siemens 3T MR. The scans comprise a total of $35$ tumor abnormalities across all subjects along with surgical pathology as the ground truth for each of them. By considering multiple disjoint tumor regions present among subjects as distinct inputs, we can increase the number of training samples.

\subsection{Training setup}
Training experiments are performed with both single and multi-modal image input combinations consecutively to perform a comparative evaluation of their classification performance. Adjustments for class imbalance are done by using weighted categorical cross-entropy based on the training distribution of labels (TN:TP balanced class weight ratios $1.95:0.65$). For our cross-validation approach, we select Leave-one-out cross-validation (LOOCV), given the low sample size to perform a less biased and thorough measure of test metrics while utilizing most of the training data. Training is performed with the Adam optimizer and learning rate $1e^{-5}$ with a batch size of $2$, consistently across all iterations for each experiment on a single P100 GPU with 16GB VRAM.
For evaluation, we compute the accuracy, recall, and precision over the entire set of test predictions ($35$ test samples) across all leave-one-out iterations.

\subsection{Results Analysis and Discussion}
%%%
\textbf{Table \ref{table:comp}} showcases classification metrics of our model, compared with the baseline experiments. Our metrics clearly show that the combination of anatomical differences ingrained in conventional MRI and metabolic differences captured by parametric PET Ki maps yields fairly high classification performance (best accuracy of $0.74$) compared to other modalities. Moreover, this also shows that the dual-encoder architecture, which encodes image modalities independently before fusing features outperforms compared to the simultaneous dual-channel and single-encoder architectures across all test metrics. Comparing image modalities independently trained with single encoder CNNs, we also observe that Ki alone performs better than MR and SUV with accuracies $0.71$, $0.65$, and $0.56$, respectively. Although we do not see drastic improvements, there is an incremental increase in the accuracy and recall between 4-5\%. Single and dual encoder's similar accuracies can, however, be attributed to the lower sample size. 

% \vspace{8pt}

Prior works have explored radiomics feature extraction from multi-modal MRI \cite{gao2020deep} and diffusion MRI \cite{park2021differentiation} along with feature selection and oversampling methods for this task but yield limited accuracy and have not evaluated the combination of metabolic image modalities such as PET with these structural ones. Another work follows the same input function derivation methodologies to compute average tumor Ki and other kinetic rate constants for classification using linear regression models \cite{schetlick2021parametric}. However, in comparison, not only do we utilize image features directly and develop more complex deep learning-based CNN approach, which can scale better with more data, but also utilize multi-modal combinations involving MRI and perform evaluation against static PET SUV maps for classification.

\section{Conclusion}
Prediction accuracy may still be limited due to the low sample size and the resulting class imbalance. Moreover, the application of data augmentation e.g. through affine image transformations is not possible because of the high heterogeneity in structure among these tumors, and applying random transformations without a better understanding of differentiating tumor features, could alter the associated class. To overcome these problems and train robust image-based classification models that could be deployed in clinical space, future improvements could include better pre-processing and feature extraction pipelines and machine learning techniques developed for small-scale datasets. Despite the low sample size while accounting for class imbalance, the current evaluation metrics elucidate that parametric PET Ki which models underlying glucose transport into the tumors performs better than static PET SUV for the classification of progressing versus necrotic tumor volumes and hence, serves as a useful addition comprising metabolic information to structural differences incorporated within MRI. For future work, we will consider incorporating deep learning-based automatic segmentation, for an end-to-end classification model. 

\section{Compliance with Ethical Considerations}
The dynamic FDG PET human imaging studies were approved by the Institutional Review Board (IRB) at the University of Virginia under protocol numbers IRB-HSR \# 17556 and IRB-HSR \#  190096. 

\section{Acknowledgements}
The authors acknowledge grant funding from the University of Virginia Brain Institute.
% Below is an example of how to insert images. Delete the ``\vspace'' line,
% uncomment the preceding line ``\centerline...'' and replace ``imageX.ps''
% with a suitable PostScript file name.
% -------------------------------------------------------------------------
% \begin{figure}[htb]

% \begin{minipage}[b]{1.0\linewidth}
%   \centering
%   \centerline{\includegraphics[width=8.5cm]{example-image}}
% %  \vspace{2.0cm}
%   \centerline{(a) Result 1}\medskip
% \end{minipage}
% %
% \begin{minipage}[b]{.48\linewidth}
%   \centering
%   \centerline{\includegraphics[width=4.0cm]{example-image}}
% %  \vspace{1.5cm}
%   \centerline{(b) Results 3}\medskip
% \end{minipage}
% \hfill
% \begin{minipage}[b]{0.48\linewidth}
%   \centering
%   \centerline{\includegraphics[width=4.0cm]{example-image}}
% %  \vspace{1.5cm}
%   \centerline{(c) Result 4}\medskip
% \end{minipage}
% %
% \caption{Example of placing a figure with experimental results.}
% \label{fig:res}
% %
% \end{figure}

\bibliographystyle{IEEEbib}

\bibliography{strings,refs}

\end{document}